\documentclass[citeauthoryear]{llncs}
\usepackage{llncsdoc}
\usepackage{graphicx}
\usepackage{epstopdf}
\usepackage{amsmath}
\begin{document}
\title{Interpolating  between Random Walks and Shortest Paths: a Path Functional Approach}

\author{Fran\c{c}ois Bavaud and Guillaume Guex \thanks{The specific remarks of two anonymous reviewers are gratefully acknowledged}}

\institute{Department of Computer Science and Mathematical Methods \\ Department of Geography\\  University of Lausanne,  Switzerland}

\maketitle

\begin{abstract}
General models of network navigation must contain a deterministic or drift component, encouraging the agent to follow routes of least cost, as well as a random or diffusive component, enabling free wandering. This paper proposes a thermodynamic formalism involving two path functionals, namely an energy functional governing the drift and an entropy functional governing the diffusion.
A freely adjustable parameter, the temperature, arbitrates between the conflicting objectives of minimising
travel costs and maximising spatial exploration. The theory is illustrated on various graphs and various temperatures.  The resulting optimal paths, together with presumably new  associated edges and nodes centrality indices, are analytically and numerically investigated.
\end{abstract}

\section{Introduction}
\vspace{-13.3cm}\textsf{\footnotesize Bavaud, F., Guex, G.: {\em Interpolating  between Random Walks and Shortest Paths:  a Path Functional Approach}. In: Aberer, K.  et al. (Eds.) SocInfo 2012, LNCS 7710, pp. 68--81. Springer, Heidelberg (2012)}\vspace{12cm}

Consider a network together with an agent wishing to move (or wishing to move goods, money, information, etc.) from source node $s$ to target node $t$. The agent seeks to minimise the total cost or duration of the move, but the ideal path may be difficult to realise exactly, in absence of perfect information about the network.

The above context is common to many behavorial and decision contexts, among which ``small-word" social communications (Travers and Milgram 1969), spatial navigation (e.g. Farnsworth and Beecham 1999), routing strategy on internet networks (e.g. Zhou 2008, Dubois-Ferri\`ere et al. 2011),  and  several others (e.g. Borgatti 2005; Newman 2005).

Trajectories can be coded, generally non-univocally, by
$X=(x_{ij})$ where $x_{ij}=\mbox{``number of  direct transitions from node $i$ to node $j$"}$. The use of the flow matrix $X$ is central in Operational Research (e.g. Ahuja et al. 1993) and Markov Chains theory (e.g. Kemeny  and Snell 1976); four optimal $st$-paths have in particular been extensively analysed {\em separately} in the litterature, namely the shortest-path, the random walk, the maximum flow (Freeman et al.  1991) and the electrical current (Kirchhoff  1850;  Newman 2005; Brandes and Fleischer 2005).

This paper investigates the properties of $st$-paths  resulting from the minimisation of a {\em free energy functional} $F(X)$, over the set $X\in {\cal X}$ of admissible solutions. $F(X)$ contains a resistance component privileging  shortest paths, and an entropy component favouring random walks. The conflict is arbitrered by a continuous parameter $T\ge0$, the {\em temperature} (or its {\em inverse} $\beta:=1/T$), and results in an analytically solvable  unique  optimum {\em continuously interpolating} between shortest-paths and random walks. See Yen et al. (2008) and  Saerens and al. (2009) for a  close proposal, yet distinct  in its   implementation.

Section \ref{defsol} introduces the formalism, in particular the  {\em energy functional} (based upon an edge resistance matrix $R$, symmetrical or not) and the {\em entropy functional}  (based upon a Markov transition matrix $W$, reversible or not, related to $R$ {\em or not}). Section \ref{algebraic} provides the analytic form of the unique solution minimising the free energy. Section \ref{evb} proposes the definition of edge and vertex betweenness centrality indices directly based upon the flow $X$. They are
illustrated in sections \ref{ics1} and \ref{ics2} for various network geometries  at various temperatures.

\section{Definitions and solutions}
\label{defsol}

\subsection{Admissible paths}
\label{admpath}
Consider a connected graph $G=(V,E)$  involving $n=|V|$ nodes together with two distinguished and distinct nodes, the source $s$ and target $t$.
The $st$-path or flow matrix, noted  $X^{st}=(x_{ij}^{st})$ or simply $X=(x_{ij})$, counts the number of transitions from $i$ to $j$
along  conserved  unit paths starting at  $s$,  possibly visiting $s$ again, and absorbed  at   $t$. Hence
\begin{eqnarray}
x_{ij} & \ge &0 \qquad\qquad   \quad  \mbox{positivity} \label{pos}\\
x_{i\bullet}-x_{\bullet i} & = & \delta_{is}-\delta_{it}  \qquad\mbox{unit flow conservation}
\label{cons} \end{eqnarray}
where $\delta_{ij}$ is the Kronecker delta, the components of the identity matrix. Here and in the sequel, $\bullet$ denotes the summation over the values of the replaced index, as in
$x_{i\bullet}=\sum_{j=1}^n x_{ij}$. In particular, $x_{s\bullet}=x_{\bullet s}+1$. Also,
\begin{equation}
\label{abs}
x_{t\bullet}=0\qquad\qquad \mbox{absorbtion at $t$}
\end{equation}
entailing $x_{tj}=0$ for all $j$, and $x_{\bullet t}=1$.
Normalisation  (\ref{cons}) can be extended to {\em valued flows}
\begin{equation}
\label{mcons}
x_{i\bullet}-x_{\bullet i}=v(\delta_{is}-\delta_{it})\qquad\mbox{conservation for valued flow}
\end{equation}
where $v\ge0$, the amount sent through  the network,
is the {\em value} of the flow. Further familiar  constraints consist of
\begin{eqnarray}
x_{ij} & \le & c_{ij}\qquad\qquad\qquad\mbox{capacity, where $c_{ij}\ge0$}\label{cap} \\
x_{ij} & \ge & b_{ij}\qquad\qquad\qquad\mbox{minimum  flow requirement, $b_{ij}\ge0$}\label{cap2} \\
x_{\bullet j_0} & = & 0 \qquad\qquad\quad\qquad\mbox{forbidden node $j_0$}\label{taboon}\\
x_{i_0j_0} & = & 0 \qquad\qquad\quad\qquad\mbox{forbidden arc $(i_0j_0)$}\enspace.\label{tabooe}
\end{eqnarray}

\subsection{Mixtures and convexity}
Any of the above constraints  (\ref{pos}) to (\ref{tabooe}) or combinations thereof defines a {\em convex} set ${\cal X}$ of admissible $st$-paths: if $X$ and $Y$ are admissible, so
is their {\em mixture} $\alpha X +(1-\alpha)Y$ for $\alpha\in [0,1]$. Mixture of paths are generally non-integer, and can be given a probabilistic interpretation, as in
\begin{enumerate}
\item[$\bullet$] $x_{\bullet\bullet}$ = ``average time (number of transitions) for  transportation  from $s$ to $t$"
\item[$\bullet$]$x_{ij}/x_{i\bullet}$ = ``conditional probability to jump to $j$ from $i$".
\end{enumerate}
From now on,  one considers by default unit flows $X$, generally non-integer, obeying (\ref{pos}), (\ref{cons}) and (\ref{abs}).

\subsection{Path entropy and energy}
\label{pee}
Let $W=(w_{ij})$ denote  the $(n\times n)$ transition matrix of some irreducible Markov chain. A $st$-path constitutes a random walk (as defined  by $W$) iff $x_{ij}/x_{i\bullet}=w_{ij}$ for all visited node $i$, i.e. such that $x_{i\bullet}>0$. Random walk $st$-paths $X$ minimise the {\em entropy} functional
\begin{displaymath}
G(X):=\sum_{ij}x_{ij}\ln\frac{x_{ij}}{x_{i\bullet}w_{ij}}=\sum_i x_{i\bullet} K_i(X||W) = x_{\bullet \bullet} \sum_i \frac{x_{i\bullet}}{x_{\bullet \bullet}} K_i(X||W)
\end{displaymath}
where
$K_i(X||W):=\sum_j \frac{x_{ij}}{x_{i\bullet}}\ln\frac{x_{ij}}{x_{i\bullet}w_{ij}}\ge 0$ is the Kullback-Leibler divergence between the transition distributions $X$ and $W$ from $i$, taking on its minimum value zero iff $\frac{x_{ij}}{x_{i\bullet}}=w_{ij}$. Note $G(X)$ to be \emph{homogeneous}, that is $G(v X) = v\:  G(X)$ for $v > 0$, reflecting the \emph{extensivity} of $G(X)$ in the thermodynamic sense.

\

By contrast, shortest-paths and other alternative optimal paths minimize {\em resistance} or {\em energy} functionals of the general form
\begin{displaymath}
U(X):=\sum_{ij}r_{ij}\varphi(x_{ij})
\end{displaymath}
where $r_{ij}>0$ represent a cost or resistance associated to the directed arc $ij$, and $\varphi(x)$ is a smooth non-decreasing function with $\varphi(0)=0$. In particular, minimizing $U(X)$ yields
\begin{enumerate}
\item[$\bullet$] $st$-shortest paths   for the choice $\varphi(x)=x$, where $r_{ij}$ is the length of the arc $ij$
\item[$\bullet$] $st$-electric currents  from $s$ to $t$  for the choice $\varphi(x)=x^2/2$, where $r_{ij}$ is the resistance of the conductor $ij$ (see section \ref{ltl}).
\end{enumerate}
As in Statistical Mechanics, we consider in this paper the class of admissible paths minimizing the {\em free energy}
\begin{equation}
\label{freeener}
F(X):=U(X)+T\: G(X)\enspace.
\end{equation}
Here  $T>0$ is a free parameter, the {\em  temperature}, controlling for the importance of the fluctuation around the trajectory of least resistance or energy (ground sate), realised in the
low temperature limit $T\to0$. In the high temperature limit $T\to\infty$ (or $\beta\to0$, where $\beta:=1/T$ is the {\em inverse temperature}), the path consists of a random walk from $s$ to $t$ governed by $W$. Hence, minimising the free energy  (\ref{freeener})
generates for $T>0$  {\em ``heated extensions"} of classical minimum-cost problems $\min_XU(X)$, with the production of random fluctuations around the classical, ``ground state" solution.

Derivating the free energy with respect to $x_{ij}$, and expressing the conservation constraints (\ref{cons}) through Lagrange multipliers $\{\lambda_i\}$  yields the  optimality condition
\begin{equation}
\label{opt1}
T \ln\frac{x_{ij}}{x_{i\bullet}w_{ij}}+  r_{ij}\varphi'(x_{ij})=\lambda_j-\lambda_i
\end{equation}
that is
\begin{equation}
\label{opt2}
x_{ij}=x_{i\bullet}w_{ij}\exp(-\beta  [r_{ij}\varphi'(x_{ij})+\lambda_i-\lambda_j] )\enspace.
\end{equation}
The multipliers are defined up to an additive constant  (see \ref{loglad}).
In any case, $x_{ij}=0$ when $w_{ij}=0$ or  $i=t$.

\subsection{Minimum free energy and uniqueness}
\label{mfeau}
Multiplying (\ref{opt1}) by $x_{ij}$ and summing over all arcs yields an identity involving the entropy $G(X)$ of the optimal path $X$. Substitution in the free energy together with (\ref{cons})  demonstrates in turn the identity
\begin{equation}
\label{deppp}
\min_X F(X)= \sum_{ij}r_{ij}[\varphi(x_{ij})-\varphi'(x_{ij})x_{ij}]+\lambda_t-\lambda_s\enspace.
\end{equation}
The first term is negative for $\varphi(x)$  convex, positive for $\varphi(x)$ concave, and zero for the heated  shortest-path problem $\varphi(x)=x$, for which $\min_XF(X)=\lambda_t-\lambda_s$.

Also, the entropy functional is convex, that is $G(\alpha X +(1-\alpha)Y)\le \alpha G(X)+(1-\alpha)G(Y)$ for two admissible paths $X$ and $Y$ and $0\le\alpha\le 1$. The energy $U(X)$ is convex (resp. concave) iff $\varphi(x)$ is convex (resp. concave).

When a  strictly convex functional $F(X)$ possesses a   local minimum on a convex domain ${\cal X}$, the minimum is unique.
In particular, we expect the optimal flows for $\varphi(x)=x^p$ to be unique for $p>1$, but  not anymore for  $0<p<1$, where  local minima may exist; see  Alamgir and von Luxburg  (2011) on  ``$p$-resistances".

In the shortest-path problem $p=1$, the solution is unique if $T>0$ (Section  \ref{algebraic}); when $T=0$, local minima of $U(X)$ may coexist,  yet all yielding the same   value of $U(X)$.

\subsection{Algebraic solution}
\label{algebraic}
Solving (\ref{opt2}) is best done by
considering separately the target node $t$. Define
$v_{ij}:=w_{ij}\exp(-\beta r_{ij}\varphi'(x_{ij}))$ as well as the
${(n-1)\times(n-1)}$ matrix $V=(v_{ij})_{i,j\neq t}$. Also, define the ${(n-1)}$ dimensional vectors
\begin{eqnarray}
\label{abab}
a_i:=x_{i\bullet}\: {\exp(-\beta\lambda_i)}\vert_{i\neq t} &  \qquad \qquad \qquad & b_j:={ \exp(\beta\lambda_j)}\vert_{j\neq t} \\
q_i:=v_{it}\vert_{i\neq t} & \qquad \qquad \qquad \qquad & e_j:=\delta_{js} \vert_{j\neq t} \notag
\end{eqnarray}
Summing  (\ref{opt2}) over all $i$ (for $j\neq t$, resp. $j=t$), then  over all $j$ for $i\neq t$ yields, using (\ref{cons}) and ({\ref{abs})
\begin{displaymath}
V'a=a-{\exp(-\beta\lambda_s)}\: e
\qquad \qquad
a'q={\exp(-\beta\lambda_t)}
\qquad \qquad
Vb+{\exp(\beta\lambda_t)}\: q=b
\end{displaymath}
Define the ${(n-1)\times(n-1)}$ matrix  $M=(m_{ij})$ and the ${(n-1)}$ vector $z$ as
\begin{equation}
\label{opt3}
M:=(I-V)^{-1}=I+V+V^2\ldots
\qquad\qquad\qquad \qquad
z:=Mq
\end{equation}
Then $a$ and $b$ express as
\begin{displaymath}
a_i= \exp(-\beta\lambda_s)\: m_{si}
\qquad\qquad \qquad \qquad
b_j={\exp(\beta\lambda_s)}\: \frac{z_j}{z_s}={\exp(\beta\lambda_j)}
\end{displaymath}
implying incidentally
\begin{equation}
\label{loglad}
\lambda_j   =T \ln z_j+C \stackrel{\mbox{\tiny (Section \ref{sec26})}}{=}T \ln z_j+\lambda_t 
\enspace.
\end{equation}
Finally
\begin{eqnarray}
\label{sola}
x_{i\bullet}=m_{si}\: \frac{z_i}{z_s} & \qquad\qquad\qquad & x_{ij}=m_{si}\: v_{ij}\: \frac{z_j}{z_s}\quad {(i\neq t)} \\
x_{it}=m_{si}\:  \frac{q_i}{z_s} & \qquad\qquad\qquad  & x_{\bullet\bullet}=\frac{(Mz)_s}{z_s}=\frac{(M^2q)_s}{(Mq)_s}\enspace.
\label{solb}\end{eqnarray}
In general, $V$, $M$, $q$ and  $d$ depend upon $X$. Hence (\ref{sola}) and (\ref{solb}) define a recursive system, whose fixed points may be multiple if $U(X)$ is not convex (Section \ref{mfeau}), but converging to a unique solution for $p>1$.

In the heated shortest-path case $p=1$, the above quantities are independent of $X$. Hence the solution is unique,  and particularly easy to compute in one single $O(n^3)$  step, involving   matrix  inversion,  as illustrated in Sections
\ref{ics1} and \ref{ics2}.

\subsection{Probabilistic interpretation}
\label{sec26}
In addition to the absorbing target node $t$, let us
introduce another ``cemetery" or absorbing state $0$, and define an extended Markov chain $P$ on $n+1$ states with transition matrix
\begin{displaymath}
P=\left(\begin{array}{c||c|c|c}
		 & {\tt i\neq t,0} & {\tt \; t \;} & {\tt \; 0 \;} \\
		\hline\hline {\tt i \neq t,0} & V & q & \rho \\
		\hline {\tt t} & 0 & 1 & 0 \\
		\hline {\tt 0} & 0 & 0 & 1
		\end{array}\right)
\end{displaymath}
where $\rho_i=1-\sum_{k=1}^n v_{ik}$ is the probability of being    absorbed at 0 from $i$ in one step.

$M=(m_{ij})$ is the so-called {\em fundamental matrix} (see 
 (\ref{opt3}) and  Kemeny  and Snell 1976 p.46), whose components $m_{ij}$ give the {\em expected number of visits from $i$ to $j$}, before being eventually absorbed at 0 or $t$. Also, $z_i$ (with $i\neq t,0$) is the {\em survival probability}, that is to be, directly or indirectly, eventually  absorbed at $t$ rather than killed at $0$,  when starting  from $i$.   The higher the node survival probability, the higher the value of its Lagrange multiplier in view of (\ref{loglad}). 
 
 Extending the latter to $j=t$ entails the consistency condition $z_t=1$, making  $\lambda_t\ge \lambda_i$ for all  $i$. In particular, the free energy of the heated shortest-path case is, in view of  (\ref{deppp}), 
\begin{displaymath}
F(X^{st})=-T \ln z_s(T)\enspace ,
\end{displaymath}
increasing (super-linearly in $T$) with the risk of being absorbed at $0$ from $s$.

\subsection{High-temperature limit}
The energy term  in (\ref{freeener}) plays no role anymore  in the limit $T\to \infty$ (that is $\beta\to0$), and so does the  absorbing state $0$ above in view of $\rho_i=0$. In particular, $z_i\equiv 1$ and $x_{ij}^{st}=m_{si}w_{ij}$ for $i\neq t$.

Also, $x^{st}_{\bullet\bullet}$ is the expected number of transitions needed to reach $t$ from $s$. The {\em commute time distance} or {\em resistance distance} $x^{st}_{\bullet\bullet}+x^{ts}_{\bullet\bullet}$ is  known to represent a {\em squared Euclidean distance} between states $s$ and $t$: see e.g. Fouss et al. 2007, and references therein; see also 
Yen et al. (2008) and Chebotarev (2010) for further studies on resistance and shortest-path {\em distances}.
\subsection{Low-temperature limit}
\label{ltl}
Equations (\ref{opt2}), (\ref{sola})  and (\ref{solb}) show the
positivity condition   $x_{ij}\ge0$ to be  automatically satisfied,
thanks  to the entropy term $G(X)$.  However,  the latter disappears in the  limit $T\to 0$, where one faces the difficulty that  the optimality condition
(\ref{opt1})  $r_{ij}\varphi'(x_{ij})=\lambda_j-\lambda_i$  is still justified only if
$x_{ij}$ is freely adjustable, that is if $x_{ij}>0$.

For the $st$-shortest path problem $\varphi(x)=x$, one gets, assuming the solution to be unique,  the well-known characterisation (see e.g. Ahuja et al. (1993) p.107):
\begin{displaymath}
\left\{\begin{array}{cc}
r_{ij}=\lambda_j-\lambda_i      & \qquad \qquad \mbox{if  } x_{ij}>0  \\
r_{ij}>\lambda_j-\lambda_i      &  \qquad \qquad   \mbox{if  } x_{ij}=0
\end{array}\right.
\end{displaymath}
occurring in the dual formulation of the $st$-shortest path problem, namely {\em ``maximize $\lambda_t-\lambda_s$ subject to
$\lambda_j-\lambda_i\le r_{ij}$ for all $i,j$"}. Here $\lambda_i$ is the shortest-path distance from $s$ to $i$.

For the $st$-electrical circuit  problem $\varphi(x)=x^2/2$, one gets $r_{ij}x_{ij}=\lambda_j-\lambda_i$ if $x_{ij}>0$, in which case $x_{ji}>0$ cannot hold in view of the positivity of the resistances, thus forcing  $x_{ji}=0$. Hence
\begin{displaymath}
\left\{\begin{array}{lc}
x_{ij}=\frac{\lambda_j-\lambda_i}{r_{ij}}>0     &  \qquad\mbox{if } \lambda_j>\lambda_i \\
x_{ij}=0 &  \qquad\mbox{otherwise} \end{array}\right.
\end{displaymath}
expressing {\em Ohm's law} for the currrent intensity $x_{ij}$ (Kirchhoff 1850), where $\lambda_i$ is the electric potential at node $i$.

\section{Illustrations and case studies: simple flow and net flow}
\label{ics1}
Let us restrict on $st$-shortest path problems, i.e. $\varphi(x) = x$, whose free energy is homogeneous in the sense
$F(vX)=vF(X)$ where $v>0$ is the value of the flow in (\ref{mcons}).

Graphs are defined by a $n\times n$ Markov transition matrix  $W$ together with a
$n\times n$   positive resistance matrix  $R$. Fixing in addition $s$, $t$  and
$\beta$,  yields an unique \emph{simple flow} $x_{ij}^{st}$, computable for any $W$ (reversible or not)  and any
$R$ (symmetric or not) - a fairly large  set of tractable weighted networks.

An obvious class of networks consists of binary graphs, defined by a symmetric, off-diagonal adjacency matrix, with unit resistances and uniform transitions on existing edges (i.e. a simple random walk in the sense of Bollob\'as 1998).

Such are the graphs
$A$ (Figure \ref{grapha}) and $B$ (Figure \ref{graphb})  below. Graph $C$ (Figure \ref{graphc})  penalises in addition two edges forming short-cut from the point of view of $W$, but with increased values of their resistance.

\begin{figure}[ht]
	\begin{center}
	\includegraphics[width=12cm]{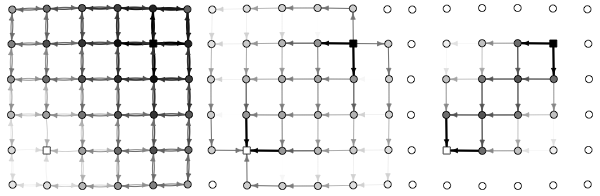}
	\caption{Graph $A$ is a square grid with uniform transitions and resistances. The resulting (high values in black, low values in light grey) simple flow $x_{ij}^{st}$ and net flow $\nu^{st}_{ij}$ from $s$ (black square) to $t$ (white square) are depicted respectively on the left and middle picture with $\beta=0$ (random walk) and on the right with $\beta=50$ (shortest-path dominance). Note the simple flow and net flow to be identical at low temperatures. }
	\label{grapha}
	\end{center}
	\end{figure}
	
\begin{figure}[ht!]
	\begin{center}
	\includegraphics[width=12cm]{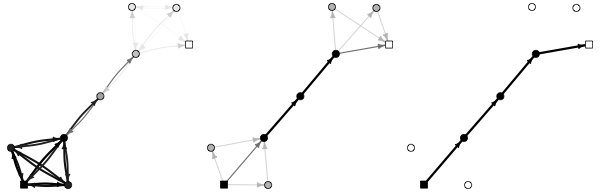}
	\caption{Graph $B$ consists of two cliques $K_4$ joined by two edges, with uniform transitions and resistances. Again, the resulting (high values in black, low values in light grey) simple flow $x_{ij}^{st}$ and net flow $\nu^{st}_{ij}$ from $s$ (black square) to $t$ (white square) are depicted respectively on the left and middle picture with $\beta=0$ (random walk) and on the right with $\beta=50$.}
	\label{graphb}
	\end{center}
\end{figure}

\begin{figure}[ht!]
	\begin{center}
	\includegraphics[width=12cm]{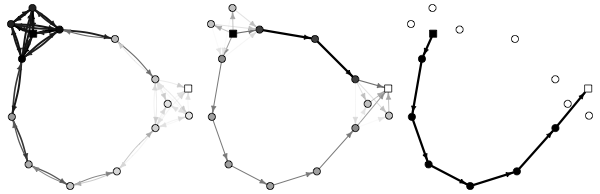}
	\caption{Graph $C$ consists of two cliques $K_5$ joined by two paths: the upper one consists of five edges, each with unit resistance, while the upper one contains  two edges, each with resistance tenfold larger. The resulting (high values in black, low values in light grey) simple flow $x_{ij}^{st}$ and net flow $\nu^{st}_{ij}$ from $s$ (black square) to $t$ (white square) are depicted respectively on the left and middle picture with $\beta=0$ (random walk) and on the right with $\beta=50$.}
	\label{graphc}
	\end{center}
\end{figure}

Among the wide variety of graphs defined by a $(W,R)$ pair, the plain graphs $A$, $B$ and $C$ primarily aim at  illustrating the basic fact  that, at high  temperature, reverberation among neighbours of the source may dramatically lengthen the shortest path - an expected phenomenon (Figure \ref{temptot}). \\

Another quantity of interest is the \emph{net flow}

\begin{equation}
\label{nf}
\nu^{st}_{ij}:=|x_{ij}^{st}-x_{ji}^{st}|
\end{equation}
discounting  ``back and forth walks" inside the same edge, as discussed by Newman (2005): as a matter of fact, the presence of such alternate moves mechanically increases the simple flow inside an edge or node, especially near the source at high temperature (Figures   \ref{grapha},  \ref{graphb} and \ref{graphc}, left), giving the false impression the behaviour is more   entropic (that is, random-walk dominated) around the source, which is erroneous.

The net flow ``filters out" reverberations and hence captures the resulting ``trend" of the agents  within their random movements, who rarely go back along the edge from where they came if there is another way; cf.  the  circulation of  ``used goods" as defined in Borgatti (2005) along  trails exempt of edges repetition. At low temperatures, the simple flow is directed in one way and hence converges to 
 the simple flow (Figures   \ref{grapha},  \ref{graphb} and \ref{graphc}, right).
 
\begin{figure}
\begin{center}
\includegraphics[width=7.5cm]{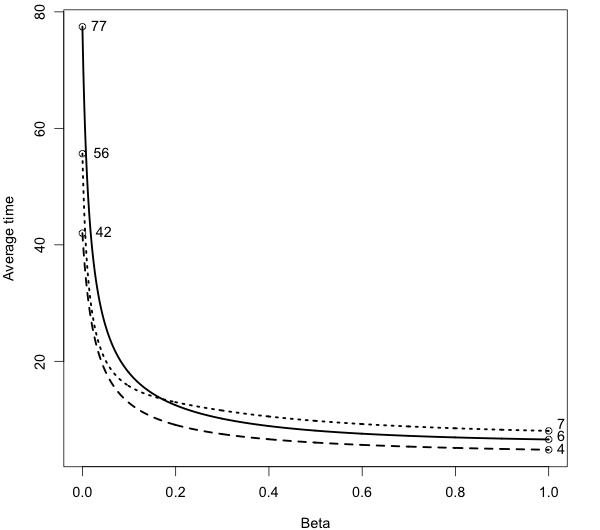}
\caption{The average time $x_{\bullet\bullet}^{st}$ to reach $t$ from $s$ is minimum for $T=0$, and decreases with the inverse temperature $\beta$. Solid line: graph $A$; Dashed line: graph $B$; Dotted line: graph $C$.}
\label{temptot}
\end{center}
\end{figure}

\section{Edge and vertex centrality betweenness}
\label{evb}
Several flow-based indices of  betweenness centrality have been proposed ever since  the shortest-path centrality pioneering proposal of Freeman (1977). In particular, random-walk centrality indices have  been discussed by  Noh  and Rieger  (2004) and Newman (2005).
In this paper, we  study the (unweighted) {\em mean flow betwenness}, defined  for edges and vertices respectively  (with complexity $O(n^5)$) as
\begin{equation}
\label{mfev}
\langle x_{ij}\rangle:=\frac{1}{n(n-1)}\sum_{s,t|s\neq t}x_{ij}^{st}
\qquad\qquad
\langle x_{i\bullet}\rangle:=\sum_j\langle x_{ij}\rangle=\langle x_{ \bullet i}\rangle
\end{equation}
where the latter identity results from the conservation condition (\ref{cons}). Definition (\ref{mfev}) is intuitive enough: an edge is central if it carries a large amount of flow  {\em on average}, that is by  considering {\em all pairs of distinct source-targets couples}, thus extending the formalism to flows without specific source or target, such as  monetary flows. 

A more formal motivation arises from sensitivity analysis, with the result \begin{displaymath}
\frac{\partial F(X(R))}{\partial r_{ij}}= \sum_{kl}\frac{\partial F(X(R))}{\partial x_{kl}(R)}
\frac{\partial x_{kl}(R)}{\partial r_{ij}}+x_{ij}(R)
=x_{ij}
\end{displaymath}
where $F(X(R))=\sum_{ij}r_{ij}\: x_{ij}(R)+TG(X(R))$ is the minimum free energy (\ref{freeener}) under the constraints of Section \ref{admpath} and
$r_{ij}$ the resistance of the edge $ij$. 

Note that $
\langle x_{\bullet\bullet}\rangle:=\sum_j\langle x_{\bullet j}\rangle$ represents the average time to go from a vertex $s$  to another vertex $t$ and to return to $s$, averaged over all distinct pairs $st$. One  can also define the {\em relative mean flow betwenness} as
\begin{displaymath}
c_{ij}:=\frac{\langle x_{ij}\rangle}{\langle x_{\bullet\bullet}\rangle}\qquad\qquad\qquad\qquad c_i:=\frac{\langle x_{i\bullet}\rangle}{\langle x_{\bullet\bullet}\rangle}
\end{displaymath}
with the property $c_{ij}\ge0$, $\sum_{ij}c_{ij}=1$ and $c_i=c_{i\bullet}=c_{\bullet i}$. \\

Another candidate for a flow-based betweenness index is  the {\em mean net flow}, again defined for edges and vertices as
\begin{equation}
\label{nmf}
\langle\nu_{ij}\rangle:=\frac{1}{n(n-1)} \sum_{s,t|s\neq t} \nu^{st}_{ij}
\qquad\qquad
\langle\nu_{i\bullet}\rangle:=\sum_j\langle\nu_{ij}\rangle=\langle\nu_{ \bullet i}\rangle
\end{equation}
Middle pictures in Figures \ref{mf1}, \ref{mf2} and  \ref{mf3} below demonstrate how the mean net flow ``substracts" the mechanical contribution  arising from back and forth walks inside the same edge, in better accordance to a common sense notion of centrality.

Also, the sensitivity of the trip duration with respect to the edge resistance
\begin{displaymath}
\sigma_{ij}:=\frac{\partial \langle x_{\bullet\bullet}(R)\rangle}{\partial r_{ij}}
\end{displaymath}
constitutes yet another candidate, amenable to analytic treatment, whose study is beyond the  size of the paper.

\section{Case studies (continued): mean flow and mean net flow}
\label{ics2}
\begin{figure}[ht!]
	\begin{center}
	\includegraphics[width=12cm]{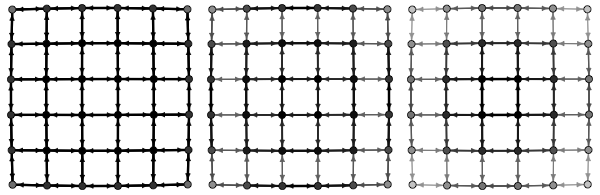}
	\caption{Graph A: mean flow $\langle x_{ij}\rangle$ and mean net flow $\langle\nu_{ij}\rangle$, with $\beta=0$ (left and middle) and $\beta=50$ (right); high values in black, low values in light grey.}
	\label{mf1}
	\end{center}
\end{figure}
\begin{figure}[ht!]
	\begin{center}
	\includegraphics[width=12cm]{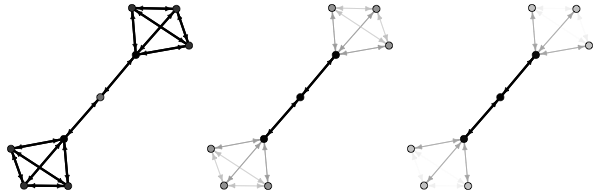}
	\caption{Graph B: mean flow $\langle x_{ij}\rangle$ and mean net flow $\langle \nu_{ij}\rangle$, with $\beta=0$ (left and middle) and $\beta=50$ (right); high values in black, low values in light grey.}
	\label{mf2}
	\end{center}
\end{figure}
\begin{figure}[ht!]
	\begin{center}
	\includegraphics[width=12cm]{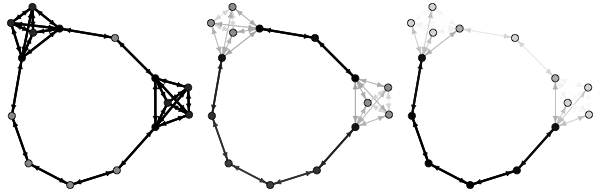}
		\caption{Graph C:  mean flow $\langle x_{ij}\rangle$ and mean net flow $\langle \nu_{ij}\rangle$, with $\beta=0$ (left and middle) and $\beta=50$ (right); high values in black, low values in light grey.}
	\label{mf3}
	\end{center}
\end{figure}

\begin{figure}[ht!]
	\begin{center}
	\includegraphics[width=12cm]{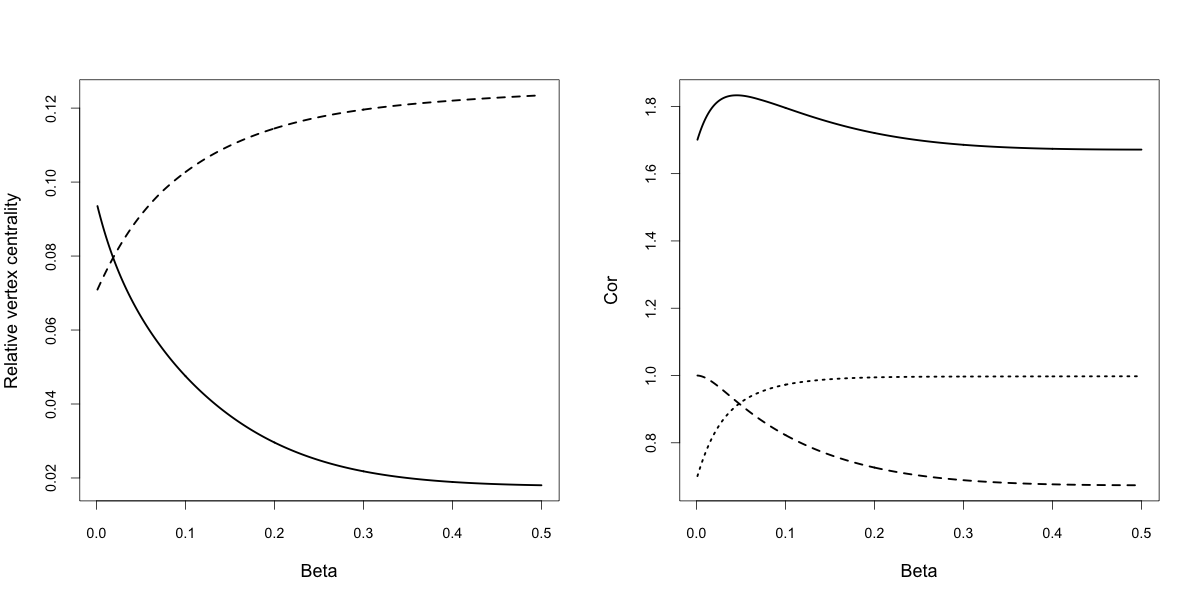}
		\caption{Left: mean net flow centrality for the vertex in the ``high-resistance path" (solid line) of network C, and for one of the nodes in the ``low-resistance path" (dashed line) of network C.
		Right: inter-nodes correlation between the mean net flow centrality with itself at $\beta=0$ (net random walk centrality; dashed line) and at  $\beta=\infty$ (shortest-path node centrality; dotted line), in function of the inverse temperature $\beta$, for graph C. The sum of the two lines (solid line) is maximum for $\beta=0.04$,  arguably indicating a transition  between an high- and a low-temperature regime.}
	\label{corcorcor}
	\end{center}
\end{figure}

Figures \ref{mf1}, \ref{mf2} and \ref{mf3} depict the mean flow betweenness and the mean net flow betweenness (\ref{mfev}) for the three graphs of Section \ref{ics1}, at high temperatures (left and middle) and low temperatures (right). Here $\langle x_{ij}\rangle=\langle x_{ji}\rangle$
due to the symmetry of $R$ and the  reversibility of $W$. Visual inspection confirms the role of the  mean flow as a  betweenness index, approaching the shortest-path betweenness at low temperatures.

At high temperatures, the mean flow $\langle x_{ij}\rangle$ turns out to be {\em constant} for all edges $ij$, a consistent observation for all ``random-walk type" networks  we have examined so far. As a consequence, the
mean flow centrality of a node $\langle x_{i\bullet}\rangle$  {\em is proportional to its degree} for $\beta\to0$, and identical to the shortest-path betweenness for $\beta\to\infty$. The former simply measures the local connectivity of the node, while the latter also takes  into account the contributions of the remote parts of the network, in particular penalising  high-resistance edges in comparison to low-resistance ones (Figure \ref{mf3}).

At low temperature, the net mean flow converges (together with the simple flow) to the shortest-path betweenness (Figures \ref{mf1}, \ref{mf2} and \ref{mf3}, right). At high temperatures, the net mean flow betweenness is large for edges connecting clusters, but, as expected, small for edges  inside clusters. Hence an original kind of centrality, the ``net random walk betweenness", differing from shortest-path and degree betweeness, can be identified (Figures \ref{mf1}, \ref{mf2} and \ref{mf3}, middle). As suggested in Figure \ref{corcorcor} (right), contributions of both origins manifest themseves in the mean flow node centrality, for intermediate values of the temperature.

\section{Conclusion}
\label{con}
The paper proposes a coherent
mechanism,  easy to implement, interpolating between shortest paths and random walks. The construction is controlled by a temperature $T$ and applies to any network endowed with a Markov transition matrix $W$ and a resistance matrix $R$. The two matrices can be related, typically as (componentwise) inverses of each other (e.g. Yen et al. 2008)  {\em or not}, in which case continuity at $T=0$ and $T=\infty$ however requires $w_{ij}>0$ whenever $r_{ij}<\infty$.

Modelling empirical  $st$-paths necessitates to define $W$ and $R$. The ``simple symmetric model", 
 namely unit resistances and uniform transitions  on existing edges  (Section \ref{ics1}) is, arguably, already  meaningful in social phenomena and otherwise. 
 For more elaborated applications,  one can  consider a possible  model of tourist paths exploring Kobe (Iryio et al. 2012),  consisting in  choosing  street directions as $W$ with a bias towards ``pleasant" street segments identified by low entries in $R$. Or  the situation where a person at $s$ wishes to be introduced to another person at $t$, by moving  over an existing social network (defined by $W$) of  friends, friends of friends, etc., where  the resistance $r_{ij}$ can express  the  difficulty  that  actor  $i$ introduces the person to actor $j$. One can also consider general situations
where $W$ expresses an average motion, a  mass  circulation, and $R$ captures an individual specific shift,  biased  towards preferentially reaching a peculiar outcome $t$, such as  a specific location, or an a-spatial goal such as  fortune, power, marriage, safety, etc. 

By contrast, the construction seems little adapted to the simulation of replicant agents (such as viruses, gossip or e-mails)    violating in general the flow conservation condition (\ref{cons}). 


The paper has defined and investigated a variety of centrality indices for edges and nodes. In particular, the mean flow betweenness interpolates between degree centrality and shortest-path centrality for nodes. Regarding edges, the mean net flow embodies various measures ranging from  simple random-walk betweenness (as defined in Newman 2005)   to shortest-path betweenness, again. The average time needed to attain  another node, respectively being attained from another node \begin{displaymath}
T^{\mbox{\tiny out}}_s:=\frac{1}{n-1}\sum_{t\: |\: t\neq s}  x_{\bullet\bullet}^{st}
\qquad\qquad\qquad
T^{\mbox{\tiny in}}_t:=\frac{1}{n-1}\sum_{s\: |\: s\neq t}  x_{\bullet\bullet}^{st}
\end{displaymath}
constitute alternative   centrality indices, generalising  Freeman's {\em closeness centrality} (Freeman 1979), incorporating  a drift component when $T > 0$.

\

Maximum-likelihood  type arguments, necessitating a probabilistic framework  not exposed here, suggest for $W$ and $R$ fixed
the estimation rule for $T$
\begin{displaymath}
U(X^{st})=U(X^{st}(T))
\end{displaymath}
where $U$ is the energy functional in Section \ref{pee}. Here $X^{st}$ is the observed, empirical path, and $X^{st}(T)$ is the optimal path (\ref{sola}, \ref{solb}) at temperature $T$. Alternatively, $T$ could be calibrated from the observed total time,  using Figure \ref{temptot} as an abacus.

\end{document}